# Formulations and Approximations of the Branch Flow Model for Mesh Power Networks

Zhao Yuan, *Member, IEEE*

*Abstract*—The formulations and approximations of the branch flow model for mesh power networks (Mesh-BranchFlow) are given in this paper. Using different sets of the power flow equations, six formats of the exact Mesh-BranchFlow model are listed. These six formats are mathematically equivalent with each other. Linear approximation and second-order cone programming (SOCP) are then used to derive the six formats of the convex Mesh-BranchFlow model. The branch ampacity constraints considering the shunt conductance and capacitance of the transmission line Π-model are derived. The key foundation of deriving the ampacity constraints is the correct interpretation of the physical meaning of the transmission line Π-model. An exact linear expression of the ampacity constraints of the power loss variable is derived. The applications of the Mesh-BranchFlow model in deriving twelve formats of the exact optimal power flow (OPF) model and twelve formats of the approximate OPF model are formulated and analyzed. Using the Julia programming language, the extensive numerical investigations of all formats of the OPF models show the accuracy and computational efficiency of the Mesh-BranchFlow model. A penalty function based approximation gap reduction method is finally proposed and numerically validated to improve the AC-feasibility of the approximate Mesh-BranchFlow model.

## Nomenclature

**Sets:**
$\mathcal{N}, \mathcal{L}$    Nodes (or buses) and lines (or branches).
$\mathcal{C}, \Omega$    Cycles (or closed loops) and decision variables.

**Variables:**
$p_n, q_n$    Active and reactive power generations at node $n$.
$p_{d_n}, q_{d_n}$    Active and reactive power demands at node $n$.
$p_{s_l}, q_{s_l}$    Non-measurable sending- end active and reactive power flows for branch $l$.
$\widetilde{p}_{s_l}, \widetilde{q}_{s_l}$    Measurable sending- end active and reactive power flows for branch $l$.
$p_{r_l}, q_{r_l}$    Non-measurable receiving- end active and reactive power injection of branch $l$.
$\widetilde{p}_{r_l}, \widetilde{q}_{r_l}$    Measurable receiving- end active and reactive power injection of branch $l$.
$q_{cs_l}, q_{cr_l}$    Sending- and receiving- end shunt reactive power of branch $l$.
$i_{s_l}, i_{r_l}$    Non-measurable sending- and receiving- end current flows of branch $l$.
$\widetilde{i}_{s_l}, \widetilde{i}_{r_l}$    Measurable sending- and receiving-end current flows of branch $l$.

$p_{o_l}, q_{o_l}$    Active and reactive power losses of branch $l$.
$v_n, V_n$    Phase-to-ground voltage magnitude and voltage square at node $n$.
$v_{s_l}, v_{r_l}$    Sending- and receiving- end phase-to-ground voltage of branch $l$.
$V_{s_l}, V_{r_l}$    Sending- and receiving- end phase-to-ground voltage square of branch $l$.
$\theta_n$    Phase-to-ground voltage phase angle at node $n$.
$\theta_l$    Phase angle difference between the sending-end and the receiving-end voltage of branch $l$.
$K^p_{o_l}, K^q_{o_l}$    Equivalent ampacity constraint for active power loss and reactive power loss.

**Parameters:**
$A^+_{nl}, A^-_{nl}$    Node to line incidence matrix.
$X_l, R_l$    Longitudinal reactance and resistance of branch $l$ modelled as a passive Π-model.
$G_n, B_n$    Shunt conductance and susceptance of node $n$.
$B_{s_l}, B_{r_l}$    Sending- and receiving- end shunt susceptance of branch $l$.
$\widetilde{K}_l, K_l$    Actual and approximate ampacity of branch $l$.
$v_n^{min}, v_n^{max}$    Lower and upper bounds of $v_n$.
$p_n^{min}, p_n^{max}$    Lower and upper bounds of $p_n$.
$q_n^{min}, q_n^{max}$    Lower and upper bounds of $q_n$.
$\theta_n^{min}, \theta_n^{max}$    Lower and upper bounds of $\theta_n$.
$\theta_l^{min}, \theta_l^{max}$    Lower and upper bounds of $\theta_l$.
$p_{d_n}, q_{d_n}$    Active and reactive power loads of node $n$.
$\alpha_n, \beta_n, \gamma_n$    Cost parameters of active power generation.
$\xi$    Penalty coefficient.

## I. Introduction

Efficient operations of power systems rely largely on the accurate modeling of power networks and the optimal solutions of the models [1]. Since the first formulation in the year of 1962, the optimal power flow (OPF) model has been investigated in enormous aspects and applied in many areas including operations, plannings, controls and market clearings [2]–[6]. [7] reports that huge economic benefits, in the scale of billions of US dollars, can be achieved for the global power industry by improving the accuracy or solution quality of the OPF model. A good summary of the traditional polar power-voltage, rectangular power-voltage and rectangular current-voltage formulations of the OPF model can be found from [7]. The large-scale integration of renewable energy resources (RERs) and the growing penetration of distributed energy resources (DERs) are pushing the power system operators to deploy the OPF model with more robust and powerful performance [8], [9]. The recent developments of OPF modeling approaches include second-order cone programming (SOCP), semi-definite programming (SDP) and polynomial

Zhao Yuan is with the Electrical Power Systems Laboratory (EPS-Lab), Department of Electrical and Computer Engineering, University of Iceland, Iceland, Email: zhaoyuan@hi.is, zhaoyuan.epslab@gmail.com.

optimization [10]–[12]. These approaches convexify the original nonconvex OPF model and thus are useful to find the global optimal solutions which are better than the local optimal solutions obtained from directly solving the original nonconvex OPF model. Compared with the local optimal solutions, the global optimal solutions can provide higher economic gains or engineering benefits. Extensive research efforts have been put to find or prove the conditions of the exactness of the SOCP based OPF models [13], to ensure the rank-1 solution of the SDP based OPF models [14] and to improve the computational efficiency of the polynomial optimization based OPF models [15]. The SOCP based OPF models feature in better computational efficiency compared with the SDP or polynomial optimization based OPF models. This is majorly because the number of variables and constraints of the SOCP based OPF models is less than the ones of the SDP or polynomial optimization based OPF models.

The dist-flow branch equations are firstly proposed in [16] to optimally size the capacitors in distribution networks. This formulation is valid only for radial power networks since no voltage phase angle constraints are considered. The voltage phase angle constraints are necessary for mesh power networks according to the Kirchhoff's laws for AC circuits. [17], [18] reformulate the dist-flow equations in [16] and denote the derived model as branch flow model. For mesh power networks, it is proved that the branch flow model in [17], [18] is valid if there are no upper bounds for the power loads. Though this condition of the power loads to validate the branch flow model in [17], [18] for mesh power networks is not realistic, this work has inspired a vast amount of research efforts to the branch flow model and its applications for example in the multi-period optimal gas-power flow (OGPF) problem and in the unbalanced three-phase distribution network context [19], [20]. Authors in [17], [18] also use SOCP to derive a convex relaxation of the branch flow model. It is shown in [21]–[23] that the ampacity constraint is not fully addressed in the branch flow model derived in [17], [18]. Authors in [21]–[23] point out that the longitude current variable used in the branch flow model in [17], [18] is not an actual measurable current according to the physical interpretation of the transmission line Π-model. As an improvement, [21]–[23] formulate an exact optimal flow (OPF) model for radial power networks. The branch flow model for mesh power networks (Mesh-BranchFlow) including the voltage phase angle constraint is firstly proposed in [24]. This Mesh-BranchFlow model is then extended, reformulated, and applied in OPF, distribution locational marginal pricing (DLMP), coordination of transmission system operator (TSO) and distribution system operator (DSO), distributed economic dispatch and super grid coordination [25]–[29]. [26] also proposes a sequential programming method to tighten the relaxation or approximation gap of the Mesh-BranchFlow model. The recent work [30] proves rigorously the relaxation property and accuracy of the reformulated convex Mesh-BranchFlow model. This paper extends the work in [30] to give more formats of the Mesh-BranchFlow model and to derive the transmission line ampacity constraint in a more accurate way.

In this paper, a comprehensive investigation of the six equivalent formats of the exact Mesh-BranchFlow model and the six formats of the approximate Mesh-BranchFlow model is conducted. The work in [30] is extended to derive the ampacity constraints considering both shunt conductive and capacitive components of the transmission line Π-model. Taking the derived ampacity constraints into account, twelve formats of the exact OPF model and twelve formats of the approximate OPF model are formulated based on the Mesh-BranchFlow equations. All the formats of the OPF model are implemented in Julia programming language and the JuMP optimization modeling package [31], [32]. A numerical investigation is then conducted through IEEE test cases. To improve the accuracy of the approximate Mesh-BranchFlow model, a penalty function based method is proposed and numerically validated. The rest of this paper is organized as follows. Section II gives the mathematical formulations of (1) the exact Mesh-BranchFlow model; (2) the approximate Mesh-BranchFlow model; (3) the branch ampacity constraint. Section III formulates the exact OPF model, the approximate OPF model and provides the numerical validations. Section IV proposes the penalty function based method to reduce the approximation gap of the approximate Mesh-BranchFlow model and numerically validates this method. Section V concludes.

## II. MESH-BRANCHFLOW MODEL

### A. Exact Model

It is assumed that the three-phase power network is balanced and the considered decision variables are constrained in (1a)-(1g):

$$v_n \in (v_n^{min}, v_n^{max}) \subseteq (0.9, 1.1), \forall n \in \mathcal{N} \quad (1a)$$

$$\theta_n \in (\theta_n^{min}, \theta_n^{max}) \subseteq (0, 2\pi), \forall n \in \mathcal{N} \quad (1b)$$

$$\theta_l = \theta_{s_l} - \theta_{r_l} \in (\theta_l^{min}, \theta_l^{max}) \subseteq (-\frac{\pi}{2}, \frac{\pi}{2}), \forall l \in \mathcal{L} \quad (1c)$$

$$p_n \in (p_n^{min}, p_n^{max}) \subset \mathbb{R}^+, \forall n \in \mathcal{N} \quad (1d)$$

$$q_n \in (q_n^{min}, q_n^{max}) \subset \mathbb{R}, \forall n \in \mathcal{N} \quad (1e)$$

$$p_{d_n} \in (p_{d_n}^{min}, p_{d_n}^{max}) \subset \mathbb{R}^+, \forall n \in \mathcal{N} \quad (1f)$$

$$q_{d_n} \in (q_{d_n}^{min}, q_{d_n}^{max}) \subset \mathbb{R}^+, \forall n \in \mathcal{N} \quad (1g)$$

Where $n \in \mathcal{N}$ is the index of nodes or buses. $l \in \mathcal{L}$ is the index of lines or branches. $v_n$ is the phase-to-ground voltage magnitude at node $n$. $\theta_n$ is the voltage phase angle at node $n$. $\theta_l$ is the phase angle difference between the sending-end voltage phase angle $\theta_{s_l}$ and the receiving-end voltage phase angle $\theta_{r_l}$ of branch $l$. Note the subscripts $s$, $r$ are not indexes but only to denote the meaning of sending and receiving. $p_n, q_n$ are the active and reactive power generations at node $n$. $p_{d_n}, q_{d_n}$ are the active and reactive power demands at node $n$. Note the subscript $d$ is not an index but only to denote the meaning of demand. We take $p_{d_n}, q_{d_n}$ as variables here to consider possible demand side responses. The demands are equal to fixed values if there is no demand side response. $p_n^{min}, p_n^{max}$ are the lower and upper bounds of $p_n$. $q_n^{min}, q_n^{max}$ are the lower and upper bounds of $q_n$. $\theta_l^{min}, \theta_l^{max}$ are the lower and upper bounds of $\theta_l$. It is assumed that $(v_n^{min}, v_n^{max}) \subset (0.9, 1.1)$ in (1a) and $(\theta_l^{min}, \theta_l^{max}) \subset (-\frac{\pi}{2}, \frac{\pi}{2})$ in (1c). These assumptions are valid in power system operations under normal conditions.

The nodal power balance equations of the Mesh-BranchFlow model are formulated in (1h)-(1i):

$$p_n - p_{d_n} = \sum_l (A_{nl}^+ p_{s_l} - A_{nl}^- p_{o_l}) + G_n v_n^2, \ \forall n \in \mathcal{N} \quad (1h)$$

$$q_n - q_{d_n} = \sum_l (A_{nl}^+ q_{s_l} - A_{nl}^- q_{o_l}) - B_n v_n^2, \ \forall n \in \mathcal{N} \quad (1i)$$

Equation (1h) represents the active power balance. Equation (1i) represents the reactive power balance. $p_{s_l}, q_{s_l}$ are the power flows at the sending-end of line $l$. The subscript $_s$ here is not an index but only to denote the meaning of sending-end. $G_n, B_n$ are the shunt conductance and susceptance at node $n$. $p_{o_l}, q_{o_l}$ are the active and reactive power loss of branch $l$. Note the subscript $_o$, is not an index but only to denote the meaning of power loss. $A_{nl}^+$ and $A_{nl}^-$ are the node-to-branch incidence matrices of the power network defined as $A_{nl}^+ = 1$, $A_{nl}^- = 0$ if $n$ is the sending-end of branch $l$, and $A_{nl}^+ = -1$, $A_{nl}^- = -1$ if $n$ is the receiving-end of branch $l$. The default convention of the sending-end or receiving-end of the lines can be defined in anyways. The only difference is that, the results of the power flow variables $p_{s_l}, q_{s_l}$ from the OPF calculations are negative if the default sending-end or receiving-end are reversed. Fig. 1 illustrates this issue. Note we neglect the power loss in Fig. 1 for sake of simplicity. The default conventions are made before the OPF calculations are done. In convention 1, node $n_1$ is referred as the sending-end and node $n_2$ is referred as the receiving-end. On the contrary, in convention 2, node $n_1$ is referred as the receiving-end and node $n_2$ is referred as the sending-end. In this set-up, after the OPF calculations are done, 100 kW power flow from node $n_1$ to node $n_2$ is equivalent to -100 kW power flow from node $n_2$ to node $n_1$. So the default convention of sending-end or receiving-end does not affect the OPF results.

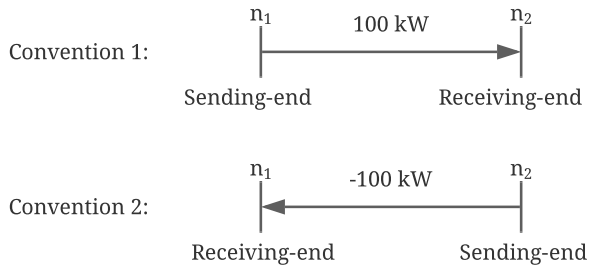

Fig. 1. Different conventions of the sending-end and receiving-end of lines.

To relate the power variables $p_{s_l}, q_{s_l}, p_{o_l}, q_{o_l}$ with the voltage variables, the voltage drop phasor of branch $l$ is used to derive the following equations (1j)-(1l):

$$v_{s_l}^2 - v_{r_l}^2 = 2R_l p_{s_l} + 2X_l q_{s_l} - R_l p_{o_l} - X_l q_{o_l}, \ \forall l \in \mathcal{L} \quad (1j)$$

$$v_{s_l} v_{r_l} sin\theta_l = X_l p_{s_l} - R_l q_{s_l}, \ \forall l \in \mathcal{L} \quad (1k)$$

$$v_{s_l}^2 - v_{s_l} v_{r_l} cos\theta_l = R_l p_{s_l} + X_l q_{s_l}, \ \forall l \in \mathcal{L} \quad (1l)$$

Where $v_{s_l}, v_{r_l}$ are the sending-end and receiving-end phase-to-ground voltage magnitudes of branch $l$. The subscripts $_s, _r$ here are not indexes but only to denote the meaning of sending-end and receiving-end. $V_{s(r)_l} = v_{s(r)_l}^2$ are voltage magnitude squares. According to the proof of Theorem 6 in the recent work [30], it is only necessary to use (1j)-(1k) or (1k)-(1l) to sufficiently express the voltage drop phasor. This reasoning is used in listing all the necessary equations of the Mesh-BranchFlow model in Table I.

The active power loss and reactive power loss $p_{o_l}, q_{o_l}$ are expressed in (1m)-(1n):

$$p_{o_l} = \frac{p_{s_l}^2 + q_{s_l}^2}{v_{s_l}^2} R_l, \ \forall l \in \mathcal{L} \quad (1m)$$

$$q_{o_l} = \frac{p_{s_l}^2 + q_{s_l}^2}{v_{s_l}^2} X_l, \ \forall l \in \mathcal{L} \quad (1n)$$

A linear relationship between $p_{o_l}$ and $p_{o_l}$ exists:

$$p_{o_l} X_l = q_{o_l} R_l, \ \forall l \in \mathcal{L} \quad (1o)$$

For mesh power networks, the sum of the phase angles of the voltage drop phasors along each closed network loop $\mathcal{C}$ should satisfy the following cyclic constraint (1p):

$$\sum_{l \in \mathcal{C}} \theta_l = 0 \ mod \ 2\pi \quad (1p)$$

Where $\mathcal{C}$ is the set of all cycles or closed loops in the meshed power networks. As it is proved and explained in [30], this constraint is implicitly satisfied if $\theta_l$ is expressed explicitly using $\theta_{s_l}, \theta_{r_l}$ in constraint (1c). So the constraint (1p) is not required in the Mesh-BranchFlow model of this paper.

The exact Mesh-BranchFlow model expressed by selecting different sets of power flow equations is summarized in Table I of this section. These six formats are mathematically equivalent. The exact Mesh-BranchFlow model is valid for both radial and mesh power networks.

TABLE I
EXACT MESH-BRANCHFLOW MODEL EXPRESSED IN DIFFERENT SETS OF EQUATIONS

| Format | Mesh-BranchFlow Equations |
|---|---|
| 1 | [(1a)-(1g), (1h)-(1k), (1m)-(1n)] |
| 2 | [(1a)-(1g), (1h)-(1k), (1m), (1o)] |
| 3 | [(1a)-(1g), (1h)-(1k), (1n)-(1o)] |
| 4 | [(1a)-(1g), (1h)-(1i), (1k)-(1l), (1m)-(1n)] |
| 5 | [(1a)-(1g), (1h)-(1i), (1k)-(1l), (1m), (1o)] |
| 6 | [(1a)-(1g), (1h)-(1i), (1k)-(1l), (1n)-(1o)] |

### B. Approximate Model

Using $V_n = v_n^2$ to replace the voltage magnitude variables, the voltage magnitude bounds (1a) can be replaced by:

$$V_n \in (V_n^{min}, V_n^{max}) \subseteq (0.81, 1.21), \ \forall n \in \mathcal{N} \quad (2a)$$

Equations (1h)-(1j) are linearized to (2b)-(2d) [30]:

$$p_n - p_{d_n} = \sum_l (A_{nl}^+ p_{s_l} - A_{nl}^- p_{o_l}) + G_n V_n, \ \forall n \in \mathcal{N} \quad (2b)$$

$$q_n - q_{d_n} = \sum_l (A_{nl}^+ q_{s_l} - A_{nl}^- q_{o_l}) - B_n V_n, \ \forall n \in \mathcal{N} \quad (2c)$$

$$V_{s_l} - V_{r_l} = 2R_l p_{s_l} + 2X_l q_{s_l} - R_l p_{o_l} - X_l q_{o_l}, \ \forall l \in \mathcal{L} \quad (2d)$$

Note the solutions of the original voltage variable $v_n$ can be obtained by $v_n = \sqrt{V_n}$ after solving the approximate Mesh-BranchFlow model.

Equation (1k) can be linearized to (2e) [30]:

$$\theta_l = X_l p_{s_l} - R_l q_{s_l}, \ \forall l \in \mathcal{L} \tag{2e}$$

It is proposed to linearize equation (1l) to (2f) in this paper as:

$$\frac{V_{s_l} - V_{r_l}}{2} = R_l p_{s_l} + X_l q_{s_l}, \ \forall l \in \mathcal{L} \tag{2f}$$

This linearization is based on the approximation $v_{s_l} v_{r_l} cos\theta_l \approx \frac{V_{s_l}+V_{r_l}}{2}$ because $v_{s_l} v_{r_l} \approx \frac{V_{s_l}+V_{r_l}}{2}$ for $v_n \in (0.9, 1.1)$ and $\theta_l \approx 0$ are valid in power system operations under normal conditions.

Using rotated second-order cone, the equations (1m)-(1n) can be approximated to (2g)-(2h):

$$p_{o_l} \geq \frac{p_{s_l}^2 + q_{s_l}^2}{V_{s_l}} R_l, \ \forall l \in \mathcal{L} \tag{2g}$$

$$q_{o_l} \geq \frac{p_{s_l}^2 + q_{s_l}^2}{V_{s_l}} X_l, \ \forall l \in \mathcal{L} \tag{2h}$$

Note equations (2d)-(2h) are convex since they are rotated cones.

The approximate Mesh-BranchFlow model expressed by selecting different sets of the linearized power flow equations is summarized in Table II of this section. These six formats are not mathematically equivalent because of the linearizations and approximations of different power flow equations. They are approximate to each other. The approximate Mesh-BranchFlow model is valid for both radial and mesh power networks.

TABLE II
APPROXIMATE MESH-BRANCHFLOW MODEL EXPRESSED IN DIFFERENT SETS OF EQUATIONS

| Format | Mesh-BranchFlow Equations |
|---|---|
| 1 | [(1b)-(1g), (2a)-(2e), (2g)-(2h)] |
| 2 | [(1b)-(1g), (2a)-(2e), (1o), (2g)] |
| 3 | [(1b)-(1g), (2a)-(2e), (1o), (2h)] |
| 4 | [(1b)-(1g), (2a)-(2c), (2e)-(2f), (2g)-(2h)] |
| 5 | [(1b)-(1g), (2a)-(2c), (1o), (2e)-(2f), (2g)] |
| 6 | [(1b)-(1g), (2a)-(2c), (1o), (2e)-(2f), (2h)] |

### C. Ampacity Constraint

The ampacity constraint of the transmission or distribution line is a very important constraint to avoid over-loading of the corresponding transmission or distribution line. It is important to emphasize the significance of the ampacity constraint considering that several big black-outs such as the Northeast blackout of 2003 in the United States are caused by the over-loading of transmission lines [33]. This paper extends the recent work [30] to consider both shunt conductance and capacitance of the transmission line Π-model in deriving the ampacity constraint. Though the shunt conductance is very small normally, it is more accurate to quantify the ampacity constraint by considering this element. The key point here is a correct interpretation of the physical meaning of the transmission line Π-model i.e. the line-to-ground shunt capacitance and conductance are actually distributed and there is a difference between the actual measurable power flow variables $\widetilde{p}_{s_l}, \widetilde{q}_{s_l}$ and the power flow variables $p_{s_l}, q_{s_l}$ used in the Mesh-BranchFlow equations. Details about this difference are analyzed in the recent work [30].

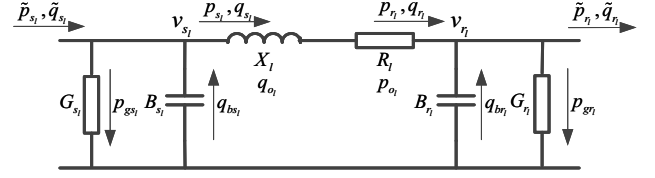

Fig. 2. Transmission line Π-model considering shunt capacitance and conductance.

From the transmission line Π-model shown in Fig. 2, the following equations can be derived:

$$\widetilde{p}_{s_l} = p_{s_l} + p_{gs_l} \tag{3a}$$

$$\widetilde{q}_{s_l} = q_{s_l} - q_{bs_l} \tag{3b}$$

$$p_{gs_l} = V_{s_l} G_{s_l} \tag{3c}$$

$$q_{bs_l} = V_{s_l} B_{s_l} \tag{3d}$$

Where $p_{gs_l}$ is the active power injection to the sending-end shunt conductance of the branch $l$ (the subscript $_{gs}$ is not an index), $q_{bs_l}$ is the reactive power injection from the sending-end shunt capacitance for the branch $l$ (the subscript $_{bs}$ is not an index), $G_{s_l}$ is the shunt conductance (the subscript $_s$ is not an index), $B_{s_l}$ is the shunt susceptance (the subscript $_s$ is not an index). The branch ampacity constraint is derived in (3e):

$$\left\|\widetilde{i}_{s_l}\right\|^2 = \frac{\widetilde{p}_{s_l}^2 + \widetilde{q}_{s_l}^2}{V_{s_l}} \leq \widetilde{K}_l \tag{3e}$$

Where $\widetilde{K}_l$ is the amapcity bound of the transmission line normally provided by the transmission line manufacturer. $\widetilde{q}_{s_l}, \widetilde{p}_{s_l}, \widetilde{i}_{s_l}$ can be measured from the sending-end of the transmission line. But $q_{s_l}, p_{s_l}, i_{s_l}$ can not be measured because they are physically distributed across the transmission line. From (3a)-(3e), the gap $\Delta^2 I$ between the current amplitude square $\|i_{s_l}\|^2$ and the measurable current amplitude square $\left\|\widetilde{i}_{s_l}\right\|^2$ in (3f) can be formulated as:

$$\Delta^2 I = \|i_{s_l}\|^2 - \left\|\widetilde{i}_{s_l}\right\|^2 = \frac{-q_{bs_l}^2 + 2q_{s_l}q_{bs_l} - p_{gs_l}^2 - 2p_{s_l}p_{gs_l}}{V_{s_l}}$$

$$= \frac{-V_{s_l}^2 B_{s_l}^2 + 2q_{s_l}V_{s_l}B_{s_l}}{V_{s_l}} + \frac{-V_{s_l}^2 G_{s_l}^2 - 2p_{s_l}V_{s_l}G_{s_l}}{V_{s_l}}$$

$$= -V_{s_l}B_{s_l}^2 + 2q_{s_l}B_{s_l} - V_{s_l}G_{s_l}^2 - 2p_{s_l}G_{s_l} \tag{3f}$$

The branch ampacity constraint (3e) is equivalent to (3g):

$$i_{s_l}^2 = \frac{p_{s_l}^2 + q_{s_l}^2}{V_{s_l}} \leq K_l = \widetilde{K}_l + \Delta^2 I \tag{3g}$$

The upper bounds of active power and reactive power loss $K_{o_l}^p, K_{o_l}^q$ can be quantified as (3h)-(3i):

$$K_{o_l}^p = K_l R_l = (\widetilde{K}_l + \Delta^2 I) R_l$$
$$= (\widetilde{K}_l - V_{s_l}B_{s_l}^2 + 2q_{s_l}B_{s_l} - V_{s_l}G_{s_l}^2 - 2p_{s_l}G_{s_l})R_l \tag{3h}$$

$$K_{o_l}^q = K_l X_l = (\widetilde{K}_l + \Delta^2 I) X_l$$
$$= (\widetilde{K}_l - V_{s_l}B_{s_l}^2 + 2q_{s_l}B_{s_l} - V_{s_l}G_{s_l}^2 - 2p_{s_l}G_{s_l})X_l \tag{3i}$$

TABLE III
FORMATS OF THE EXACT OPF MODEL

| Format | Exact OPF Model |
|---|---|
| 1 | $\operatorname*{argmin}_{\Omega} f(\Omega) := \{\Omega \in [(1a)-(1g),(1h)-(1k),(1m)-(1n),(3h),(3j)]\}$ |
| 2 | $\operatorname*{argmin}_{\Omega} f(\Omega) := \{\Omega \in [(1a)-(1g),(1h)-(1k),(1m),(1o),(3h),(3j)]\}$ |
| 3 | $\operatorname*{argmin}_{\Omega} f(\Omega) := \{\Omega \in [(1a)-(1g),(1h)-(1k),(1n)-(1o),(3h),(3j)]\}$ |
| 4 | $\operatorname*{argmin}_{\Omega} f(\Omega) := \{\Omega \in [(1a)-(1g),(1h)-(1i),(1k)-(1l),(1m)-(1n),(3h),(3j)]\}$ |
| 5 | $\operatorname*{argmin}_{\Omega} f(\Omega) := \{\Omega \in [(1a)-(1g),(1h)-(1i),(1k)-(1l),(1m),(1o),(3h),(3j)]\}$ |
| 6 | $\operatorname*{argmin}_{\Omega} f(\Omega) := \{\Omega \in [(1a)-(1g),(1h)-(1i),(1k)-(1l),(1n)-(1o),(3h),(3j)]\}$ |
| 7 | $\operatorname*{argmin}_{\Omega} f(\Omega) := \{\Omega \in [(1a)-(1g),(1h)-(1k),(1m)-(1n),(3i),(3k)]\}$ |
| 8 | $\operatorname*{argmin}_{\Omega} f(\Omega) := \{\Omega \in [(1a)-(1g),(1h)-(1k),(1m),(1o),(3i),(3k)]\}$ |
| 9 | $\operatorname*{argmin}_{\Omega} f(\Omega) := \{\Omega \in [(1a)-(1g),(1h)-(1k),(1n)-(1o),(3i),(3k)]\}$ |
| 10 | $\operatorname*{argmin}_{\Omega} f(\Omega) := \{\Omega \in [(1a)-(1g),(1h)-(1i),(1k)-(1l),(1m)-(1n),(3i),(3k)]\}$ |
| 11 | $\operatorname*{argmin}_{\Omega} f(\Omega) := \{\Omega \in [(1a)-(1g),(1h)-(1i),(1k)-(1l),(1m),(1o),(3i),(3k)]\}$ |
| 12 | $\operatorname*{argmin}_{\Omega} f(\Omega) := \{\Omega \in [(1a)-(1g),(1h)-(1i),(1k)-(1l),(1n)-(1o),(3i),(3k)]\}$ |

TABLE IV
FORMATS OF THE APPROXIMATE OPF MODEL

| Format | Approximate OPF Model |
|---|---|
| 1 | $\operatorname*{argmin}_{\Omega} f(\Omega) := \{\Omega \in [(1b)-(1g),(2a)-(2e),(2g)-(2h),(3h),(3j),(4a)]\}$ |
| 2 | $\operatorname*{argmin}_{\Omega} f(\Omega) := \{\Omega \in [(1b)-(1g),(2a)-(2e),(1o),(2g),(3h),(3j),(4a)]\}$ |
| 3 | $\operatorname*{argmin}_{\Omega} f(\Omega) := \{\Omega \in [(1b)-(1g),(2a)-(2e),(1o),(2h),(3h),(3j),(4a)]\}$ |
| 4 | $\operatorname*{argmin}_{\Omega} f(\Omega) := \{\Omega \in [(1b)-(1g),(2a)-(2c),(2e)-(2f),(2g)-(2h),(3h),(3j),(4a)]\}$ |
| 5 | $\operatorname*{argmin}_{\Omega} f(\Omega) := \{\Omega \in [(1b)-(1g),(2a)-(2c),(1o),(2e)-(2f),(2g),(3h),(3j),(4a)]\}$ |
| 6 | $\operatorname*{argmin}_{\Omega} f(\Omega) := \{\Omega \in [(1b)-(1g),(2a)-(2c),(1o),(2e)-(2f),(2h),(3h),(3j),(4a)]\}$ |
| 7 | $\operatorname*{argmin}_{\Omega} f(\Omega) := \{\Omega \in [(1b)-(1g),(2a)-(2e),(2g)-(2h),(3i),(3k),(4a)]\}$ |
| 8 | $\operatorname*{argmin}_{\Omega} f(\Omega) := \{\Omega \in [(1b)-(1g),(2a)-(2e),(1o),(2g),(3i),(3k),(4a)]\}$ |
| 9 | $\operatorname*{argmin}_{\Omega} f(\Omega) := \{\Omega \in [(1b)-(1g),(2a)-(2e),(1o),(2h),(3i),(3k),(4a)]\}$ |
| 10 | $\operatorname*{argmin}_{\Omega} f(\Omega) := \{\Omega \in [(1b)-(1g),(2a)-(2c),(2e)-(2f),(2g)-(2h),(3i),(3k),(4a)]\}$ |
| 11 | $\operatorname*{argmin}_{\Omega} f(\Omega) := \{\Omega \in [(1b)-(1g),(2a)-(2c),(1o),(2e)-(2f),(2g),(3i),(3k),(4a)]\}$ |
| 12 | $\operatorname*{argmin}_{\Omega} f(\Omega) := \{\Omega \in [(1b)-(1g),(2a)-(2c),(1o),(2e)-(2f),(2h),(3i),(3k),(4a)]\}$ |

Because $G_{s_l}, B_{s_l}$ are constants, equations (3h)-(3i) are linear (the final expressions). The expression of $K^p_{o_l}, K^q_{o_l}$ from (3h)-(3i) is used in the exact and approximate Mesh-BranchFlow model. The approximate Mesh-BranchFlow model is still convex. In this way, any approximation on the branch ampacity constraint is avoided. This ampacity constraint can be used to constrain the power loss variable which equivalently constrains the capacity of the transmission lines expressed in (3j)-(3k):

$$p_{o_l} \leq K^p_{o_l} \quad (3j)$$

$$q_{o_l} \leq K^q_{o_l} \quad (3k)$$

## III. OPTIMAL POWER FLOW

The OPF problem is a fundamental mathematical optimization model used widely in power system operations. Power system operators use OPF to take decisions in the control room. The objective of the decision making can be minimizing the economic generation cost (economic dispatch), minimizing the power loss and maximizing the security margin .etc. Any decisions in operating the power network must take into account the physical laws of power flow and operational constraints. Here, the set $\Omega \subseteq \{p_n, q_n, p_{d_n}, q_{d_n}, v_n, V_n, \theta_l, p_{s_l}, q_{s_l}, p_{o_l}, q_{o_l}\}$ is used to represent the set of decision variables in expressing the OPF model. By deploying one format of the exact Mesh-BranchFlow model in Table I or the approximate Mesh-BranchFlow model in Table II as the constraints of the OPF problem, different formats of the exact OPF problem or the approximate OPF problem can be formulated. For the branch ampacity constraint, either constraints (3h)-and-(3j) or (3i)-and-(3k) described in the section II of this paper can be used. Twelve formats of the exact OPF model are listed in Table III. They are mathematically equivalent.

To improve the AC-feasibility of the approximate Mesh-BranchFlow model, the following conic constraint is proved to be a necessary condition to recover AC-feasible solution [30]:

$$V_{s_l} V_{r_l} sin^2(\theta_l^{max}) \geq \theta_l^2, \; \forall l \in \mathcal{L} \quad (4a)$$

This constraint is included in the approximate OPF model based on the approximate Mesh-BranchFlow equations. Since this constraint is conic, it is convex. Twelve formats of the approximate OPF model are listed in Table IV. They are not mathematically equivalent but are approximate to each other.

TABLE V
OBJECTIVE SOLUTION OF THE EXACT OPF MODEL [$]

| Format | case9 | IEEE14 | case30 | IEEE57 | case89pegase | IEEE118 | ACTIVSg200 | IEEE300 | ACTIVSg500 |
|---|---|---|---|---|---|---|---|---|---|
| 1 | 5296.69 | 8081.61 | 576.89 | 41738.11 | 5819.69 | 129660.63 | 27557.57 | 719732.11 | 71817.42 |
| 2 | 5314.02 | 8078.80 | 573.94 | 41698.64 | 5818.89 | 129695.30 | 27557.57 | 719377.57 | 71817.42 |
| 3 | 5296.69 | 8081.61 | 576.89 | 41738.11 | 5819.69 | 129660.63 | 27557.57 | 719732.11 | 71817.42 |
| 4 | 5296.69 | 8081.54 | 576.89 | 41737.93 | 5819.51 | 129660.54 | 27557.57 | 719732.22 | 71817.42 |
| 5 | 5296.68 | 8080.11 | 576.21 | 41696.02 | 5831.56 | 129930.62 | 27557.57 | 719396.40 | 71817.42 |
| 6 | 5296.69 | 8081.54 | 576.89 | 41737.93 | 5819.51 | 129660.54 | 27557.57 | 719731.22 | 71817.42 |
| 7 | 5296.69 | 8081.61 | 576.89 | 41738.11 | 5819.69 | 129660.63 | 27557.57 | 719732.11 | 71817.42 |
| 8 | 5296.68 | 8078.80 | 576.54 | 41698.64 | 5849.61 | 129695.30 | 27557.57 | 719377.57 | 71817.42 |
| 9 | 5296.69 | 8081.61 | 576.89 | 41738.11 | 5819.69 | 129660.63 | 27557.57 | 719732.11 | 71817.42 |
| 10 | 5296.69 | 8081.54 | 576.89 | 41737.93 | 5819.51 | 129660.54 | 27557.57 | 719731.22 | 71817.42 |
| 11 | 5309.00 | 8080.12 | 578.99 | 41696.02 | 5819.93 | 129930.62 | 27557.57 | 719396.40 | 71817.42 |
| 12 | 5296.69 | 8081.54 | 576.89 | 41737.93 | 5819.51 | 129660.54 | 27557.57 | 719731.22 | 71817.42 |
| MATPOWER | 5296.69 | 8081.53 | 576.89 | 41737.79 | 5819.81 | 129660.70 | 27557.57 | 719725.11 | 72578.30 |

TABLE VI
COMPUTATIONAL TIME OF THE EXACT OPF MODEL [S]

| Format | case9 | IEEE14 | case30 | IEEE57 | case89pegase | IEEE118 | ACTIVSg200 | IEEE300 | ACTIVSg500 |
|---|---|---|---|---|---|---|---|---|---|
| 1 | 0.03 | 0.02 | 0.05 | 0.08 | 0.33 | 0.27 | 0.63 | 0.95 | 1.67 |
| 2 | 0.16 | 0.03 | 0.08 | 0.14 | 7.27 | 1.36 | 0.59 | 2.33 | 1.64 |
| 3 | 0.05 | 0.02 | 0.06 | 0.06 | 0.33 | 0.27 | 0.83 | 0.88 | 1.64 |
| 4 | 0.03 | 0.02 | 0.06 | 0.08 | 0.27 | 0.25 | 0.66 | 0.98 | 1.77 |
| 5 | 0.33 | 0.03 | 0.31 | 0.13 | 3.56 | 0.81 | 0.86 | 1.36 | 1.72 |
| 6 | 0.03 | 0.02 | 0.06 | 0.06 | 0.36 | 0.24 | 0.86 | 0.95 | 1.70 |
| 7 | 0.05 | 0.03 | 0.09 | 0.08 | 0.22 | 0.25 | 0.59 | 0.95 | 2.66 |
| 8 | 0.94 | 0.03 | 26.28 | 0.13 | 2.80 | 1.36 | 0.56 | 2.12 | 6.36 |
| 9 | 0.05 | 0.03 | 0.06 | 0.08 | 0.20 | 0.25 | 0.55 | 0.91 | 2.94 |
| 10 | 0.05 | 0.02 | 0.06 | 0.06 | 0.22 | 0.25 | 0.61 | 1.02 | 2.77 |
| 11 | 1.61 | 0.03 | 9.00 | 0.11 | 18.17 | 0.83 | 0.55 | 1.34 | 7.36 |
| 12 | 0.05 | 0.03 | 0.06 | 0.06 | 0.22 | 0.27 | 0.56 | 0.95 | 3.33 |
| MATPOWER | 1.20 | 1.67 | 2.14 | 1.88 | 2.50 | 2.02 | 2.09 | 2.13 | 2.78 |

For the approximate OPF model, the approximation gaps of active power loss $gap_l^{po}$ and reactive power loss $gap_l^{qo}$ are defined as:

$$gap_l^{po} := p_{o_l} - \frac{p_{s_l}^2 + q_{s_l}^2}{V_{s_l}} R_l, \ \forall l \in \mathcal{L} \tag{4b}$$

$$gap_l^{qo} := q_{o_l} - \frac{p_{s_l}^2 + q_{s_l}^2}{V_{s_l}} X_l, \ \forall l \in \mathcal{L} \tag{4c}$$

The corresponding maximum approximation gaps (of active and reactive power loss) are defined as:

$$gap^{po,max} := Maximum\{gap_l^{po}, \forall l \in \mathcal{L}\} \tag{4d}$$

$$gap^{qo,max} := Maximum\{gap_l^{qo}, \forall l \in \mathcal{L}\} \tag{4e}$$

These approximation gaps are useful to quantify the AC-feasibility of the solutions from the approximate OPF model. A fully AC-feasible solution of the approximate OPF model means that $gap^{po,max} = gap^{qo,max} = 0$. When $gap^{po,max} \neq 0$ or $gap^{qo,max} \neq 0$, smaller values of $gap^{po,max}, gap^{qo,max}$ mean better solution quality in terms of AC-feasibility.

A typical example of the OPF problem to minimize a quadratic power generation cost function as the objective function $f$ is formulated in equation (4f). This formulation is also used in the OPF formulations in MATPOWER.

$$f(p_n) = \sum_n (\alpha_n p_n^2 + \beta_n p_n + \gamma_n) \tag{4f}$$

Where $\alpha_n, \beta_n, \gamma_n \geq 0$ are the cost parameters of active power generations. Numerical investigations of the OPF model using (4f) as the objective function is conducted in this section.

All the formats of the OPF model listed in Table III and Table IV are implemented in Julia programming language and the JuMP optimization modeling package. The codes are running on the 64-bit Windows 10 operating system. A personal computer with Intel-i7 3GHz CPU and 32G RAM is deployed. The IPOPT solver is used to solve the exact OPF model and the approximate OPF model in Julia [34]. Some test cases of the approximate OPF model are solved by the CPLEX solver. The power network data (including the cost parameters $\alpha_n, \beta_n, \gamma_n$) from MATPOWER is used here [35]. The evaluated power networks include case9, IEEE14, case30, IEEE57, case89pegase, IEEE118, IEEE300 and ACTIVSg500 [36]–[39]. OPF solutions from MATPOWER are used as the benchmark.

The objective solutions of the exact OPF model are listed in Table V. All the proposed twelve formats of the exact OPF model have the same or very close objective solutions compared with MATPOWER. These results show that all the proposed formats of the exact OPF model are accurate. The computational CPU time of the exact OPF model are listed in Table VI. Note the computational time for the proposed twelve formats of the OPF model include both optimization model construction time in JuMP and the solver time in IPOPT. With the increase of the network scale, the required computational CPU time increases. This is reasonable since the number of model variables and constraints increase.

The objective solutions of the approximate OPF model

TABLE VII
OBJECTIVE SOLUTION OF THE APPROXIMATE OPF MODEL [$]

| Format | case9 | IEEE14 | case30 | IEEE57 | case89pegase | IEEE118 | ACTIVSg200 | IEEE300 | ACTIVSg500 |
|---|---|---|---|---|---|---|---|---|---|
| 1 | 5296.69 | 8080.44 | 575.33 | 41734.92 | 5814.56 | 129618.44 | 27553.67 | 719596.69 | 71891.92 |
| 2 | 5296.69 | 8080.64 | 576.51 | 41728.82 | 5819.04 | 129625.35 | 27557.57 | 719548.84 | 71893.41 |
| 3 | 5296.69 | 8081.55 | 576.85 | 41735.91 | 5819.05 | 129626.18 | 27557.57 | 719699.91 | 71893.41 |
| 4 | 5296.49 | 8081.97 | 575.68 | 41726.41 | 5813.30 | 129606.68 | 27553.70 | 719710.59 | 71930.49 |
| 5 | 5315.48 | 8081.43 | 576.28 | 41719.61 | 5816.87 | 129614.02 | 27557.58 | 719456.87 | 71931.89 |
| 6 | 5315.49 | 8081.97 | 576.29 | 41726.41 | 5816.87 | 129615.15 | 27557.58 | 719824.73 | 71931.89 |
| 7 | 5296.69 | 8080.44 | 576.66 | 41734.92 | 5815.25 | 129618.44 | 27553.79 | 719596.69 | 71891.99 |
| 8 | 5296.69 | 8080.64 | 576.51 | 41728.82 | 5819.04 | 129625.35 | 27557.57 | 719548.84 | 71893.42 |
| 9 | 5296.69 | 8081.55 | 576.85 | 41735.91 | 5819.05 | 129626.18 | 27557.57 | 719699.91 | 71893.42 |
| 10 | 5315.49 | 8081.97 | 576.15 | 41726.41 | 5813.30 | 129606.68 | 27553.70 | 719710.59 | 71930.49 |
| 11 | 5315.48 | 8081.43 | 576.28 | 41719.61 | 5816.87 | 129614.02 | 27557.58 | 719456.87 | 71931.89 |
| 12 | 5315.49 | 8081.97 | 576.29 | 41726.41 | 5816.87 | 129615.15 | 27557.58 | 719824.73 | 71931.89 |

TABLE VIII
COMPUTATIONAL TIME OF THE APPROXIMATE OPF MODEL [S]

| Format | case9 | IEEE14 | case30 | IEEE57 | case89pegase | IEEE118 | ACTIVSg200 | IEEE300 | ACTIVSg500 |
|---|---|---|---|---|---|---|---|---|---|
| 1 | 0.03 | 0.05 | 0.09 | 0.16 | 0.56 | 0.48 | 0.77 | 1.49 | 2.83 |
| 2 | 0.05 | 0.05 | 0.16 | 0.31 | 1.08 | 0.80 | 0.64 | 4.06 | 2.27 |
| 3 | 0.03 | 0.03 | 0.08 | 0.11 | 0.41 | 0.33 | 0.62 | 1.38 | 2.13 |
| 4 | 0.03 | 0.05 | 0.13 | 0.16 | 1.00 | 0.44 | 0.78 | 1.56 | 2.89 |
| 5 | 0.06 | 0.06 | 0.09 | 0.28 | 0.41 | 0.94 | 0.72 | 5.67 | 2.25 |
| 6 | 0.03 | 0.03 | 0.06 | 0.14 | 0.41 | 0.38 | 0.69 | 1.75 | 2.19 |
| 7 | 0.03 | 0.05 | 0.09 | 0.16 | 0.53 | 0.45 | 0.72 | 1.77 | 3.17 |
| 8 | 0.05 | 0.05 | 0.08 | 0.28 | 0.83 | 0.80 | 0.69 | 4.00 | 4.47 |
| 9 | 0.03 | 0.03 | 0.08 | 0.11 | 0.37 | 0.33 | 0.66 | 1.49 | 3.31 |
| 10 | 0.03 | 0.05 | 0.13 | 0.14 | 0.52 | 0.42 | 0.78 | 1.75 | 2.69 |
| 11 | 0.06 | 0.05 | 0.05 | 0.26 | 1.00 | 0.91 | 0.69 | 5.42 | 2.34 |
| 12 | 0.05 | 0.03 | 0.09 | 0.11 | 0.41 | 0.34 | 0.67 | 1.88 | 2.25 |

TABLE IX
MAXIMUM APPROXIMATION GAP OF THE ACTIVE POWER LOSS OF THE APPROXIMATE OPF MODEL

| Format | case9 | IEEE14 | case30 | IEEE57 | case89pegase | IEEE118 | ACTIVSg200 | IEEE300 | ACTIVSg500 |
|---|---|---|---|---|---|---|---|---|---|
| 1 | 0.00E+00 | 0.00E+00 | 0.00E+00 | 0.00E+00 | 2.53E-11 | 0.00E+00 | 2.01E-12 | 5.15E-14 | 7.11E-14 |
| 2 | 7.70E-14 | 1.29E-10 | 5.39E-15 | 8.00E-11 | 1.85E-02 | 2.54E-12 | 3.49E-12 | 2.84E-13 | 2.95E-15 |
| 3 | 0.00E+00 | 0.00E+00 | 0.00E+00 | 0.00E+00 | 1.85E-02 | 0.00E+00 | 3.49E-12 | 3.39E-14 | 2.95E-15 |
| 4 | 0.00E+00 | 0.00E+00 | 0.00E+00 | 0.00E+00 | 2.53E-11 | 0.00E+00 | 3.42E-13 | 5.60E-14 | 7.06E-14 |
| 5 | 3.02E-13 | 1.30E-10 | 8.23E-15 | 7.85E-11 | 1.61E-02 | 2.95E-12 | 2.09E-12 | 4.38E-03 | 3.42E-14 |
| 6 | 0.00E+00 | 0.00E+00 | 0.00E+00 | 0.00E+00 | 1.61E-02 | 0.00E+00 | 2.09E-12 | 4.32E-03 | 3.42E-14 |
| 7 | 0.00E+00 | 0.00E+00 | 0.00E+00 | 0.00E+00 | 2.53E-11 | 0.00E+00 | 2.01E-12 | 5.15E-14 | 7.11E-14 |
| 8 | 7.06E-11 | 1.29E-10 | 2.14E-08 | 8.00E-11 | 1.85E-02 | 2.54E-12 | 3.49E-12 | 2.84E-13 | 2.95E-15 |
| 9 | 0.00E+00 | 0.00E+00 | 0.00E+00 | 0.00E+00 | 1.85E-02 | 0.00E+00 | 3.49E-12 | 3.39E-14 | 2.95E-15 |
| 10 | 0.00E+00 | 0.00E+00 | 0.00E+00 | 0.00E+00 | 2.53E-11 | 0.00E+00 | 2.01E-12 | 5.60E-14 | 7.06E-14 |
| 11 | 7.06E-11 | 1.30E-10 | 4.11E-09 | 7.85E-11 | 1.61E-02 | 2.95E-12 | 2.09E-12 | 4.38E-03 | 3.42E-14 |
| 12 | 0.00E+00 | 0.00E+00 | 0.00E+00 | 0.00E+00 | 1.61E-02 | 0.00E+00 | 2.09E-12 | 4.32E-03 | 3.42E-14 |

TABLE X
MAXIMUM APPROXIMATION GAP OF THE REACTIVE POWER LOSS OF THE APPROXIMATE OPF MODEL

| Format | case9 | IEEE14 | case30 | IEEE57 | case89pegase | IEEE118 | ACTIVSg200 | IEEE300 | ACTIVSg500 |
|---|---|---|---|---|---|---|---|---|---|
| 1 | 3.32E-01 | 2.43E-01 | 3.08E-01 | 1.25E-01 | 3.47E+00 | 6.02E+00 | 4.84E-01 | 3.13E+00 | 2.98E-01 |
| 2 | 2.40E-01 | 4.63E-10 | 6.14E-02 | 1.14E-01 | 1.47E+00 | 3.99E+00 | 2.55E-10 | 1.00E+00 | 3.35E-13 |
| 3 | 3.32E-01 | 4.07E-10 | 8.49E-02 | 8.30E-02 | 1.47E+00 | 4.23E+00 | 2.55E-10 | 8.14E-01 | 3.35E-13 |
| 4 | 2.07E+00 | 2.13E-01 | 2.61E-01 | 4.75E-01 | 2.15E+00 | 7.42E+00 | 3.20E-01 | 2.27E+04 | 1.01E-01 |
| 5 | 1.71E+00 | 8.65E-02 | 1.75E-01 | 1.47E-01 | 1.27E+00 | 2.96E+00 | 1.52E-10 | 1.00E+05 | 3.88E-12 |
| 6 | 2.12E+00 | 9.58E-02 | 2.18E-01 | 1.44E-01 | 1.27E+00 | 3.45E+00 | 1.52E-10 | 1.05E+04 | 3.88E-12 |
| 7 | 2.41E-01 | 2.43E-01 | 9.89E-02 | 1.25E-01 | 1.53E+00 | 6.02E+00 | 3.69E-01 | 3.13E+00 | 1.52E-01 |
| 8 | 1.41E-01 | 4.63E-10 | 1.30E-02 | 1.14E-01 | 1.44E+00 | 3.99E+00 | 2.55E-10 | 1.00E+00 | 3.35E-13 |
| 9 | 2.41E-01 | 4.07E-10 | 2.63E-02 | 8.30E-02 | 1.47E+00 | 4.23E+00 | 2.55E-10 | 8.14E-01 | 3.35E-13 |
| 10 | 1.85E-01 | 2.13E-01 | 9.69E-02 | 4.75E-01 | 2.15E+00 | 7.42E+00 | 5.42E-01 | 2.27E+04 | 1.53E-01 |
| 11 | 1.01E-01 | 8.65E-02 | 3.28E-02 | 1.47E-01 | 1.27E+00 | 2.96E+00 | 1.52E-10 | 1.00E+05 | 3.88E-12 |
| 12 | 1.85E-01 | 9.58E-02 | 9.78E-02 | 1.44E-01 | 1.27E+00 | 3.45E+00 | 1.52E-10 | 1.05E+04 | 3.88E-12 |

TABLE XI
REDUCED MAXIMUM APPROXIMATION GAP OF THE ACTIVE POWER LOSS OF THE APPROXIMATE OPF MODEL

| Format | case9 | IEEE14 | case30 | IEEE57 | case89pegase | IEEE118 | ACTIVSg200 | IEEE300 | ACTIVSg500 |
|---|---|---|---|---|---|---|---|---|---|
| 1 | 0.00E+00 | 0.00E+00 | 0.00E+00 | 0.00E+00 | 9.41E-10 | 0.00E+00 | 1.96E-12 | 3.70E-07 | 7.21E-14 |
| 2 | 3.68E-15 | 1.29E-10 | 6.60E-15 | 8.12E-11 | 2.83E-09 | 2.72E-12 | 7.77E-13 | 3.99E-08 | 5.71E-13 |
| 3 | 0.00E+00 | 0.00E+00 | 0.00E+00 | 0.00E+00 | 6.12E-12 | 0.00E+00 | 7.77E-13 | 7.34E-08 | 5.71E-13 |
| 4 | 0.00E+00 | 0.00E+00 | 0.00E+00 | 0.00E+00 | 3.11E-08 | 0.00E+00 | 1.96E-12 | 4.55E-07 | 7.17E-14 |
| 5 | 1.09E-14 | 1.30E-10 | 3.55E-15 | 7.19E-11 | 2.44E-11 | 2.72E-12 | 1.37E-12 | 6.97E-08 | 5.66E-13 |
| 6 | 0.00E+00 | 0.00E+00 | 0.00E+00 | 0.00E+00 | 6.15E-12 | 0.00E+00 | 7.80E-13 | 1.61E-03 | 5.66E-13 |
| 7 | 0.00E+00 | 0.00E+00 | 0.00E+00 | 0.00E+00 | 3.11E-08 | 0.00E+00 | 1.96E-12 | 3.70E-07 | 7.21E-14 |
| 8 | 7.08E-11 | 1.29E-10 | 1.04E-09 | 8.13E-11 | 1.16E-07 | 2.72E-12 | 1.37E-12 | 3.99E-08 | 5.71E-13 |
| 9 | 0.00E+00 | 0.00E+00 | 0.00E+00 | 0.00E+00 | 6.12E-12 | 0.00E+00 | 7.77E-13 | 7.34E-08 | 5.71E-13 |
| 10 | 0.00E+00 | 0.00E+00 | 0.00E+00 | 0.00E+00 | 1.24E-11 | 0.00E+00 | 1.96E-12 | 4.55E-07 | 7.17E-14 |
| 11 | 7.07E-11 | 1.30E-10 | 2.84E-09 | 7.19E-11 | 2.44E-11 | 2.72E-12 | 7.80E-13 | 6.97E-08 | 5.66E-13 |
| 12 | 0.00E+00 | 0.00E+00 | 0.00E+00 | 0.00E+00 | 6.15E-12 | 0.00E+00 | 7.80E-13 | 1.61E-03 | 5.66E-13 |

TABLE XII
REDUCED MAXIMUM APPROXIMATION GAP OF THE REACTIVE POWER LOSS OF THE APPROXIMATE OPF MODEL

| Format | case9 | IEEE14 | case30 | IEEE57 | case89pegase | IEEE118 | ACTIVSg200 | IEEE300 | ACTIVSg500 |
|---|---|---|---|---|---|---|---|---|---|
| 1 | 3.94E-10 | 3.62E-10 | 1.04E-10 | 7.21E-11 | 8.50E-11 | 4.38E-11 | 7.05E-11 | 4.75E-02 | 1.95E-10 |
| 2 | 6.64E-10 | 4.74E-10 | 1.76E-10 | 8.00E-12 | 4.56E-08 | 3.43E-10 | 9.84E-12 | 3.09E-02 | 9.38E-12 |
| 3 | 3.95E-10 | 4.07E-10 | 1.76E-10 | 8.94E-11 | 9.61E-11 | 3.25E-11 | 9.84E-12 | 7.53E-07 | 9.38E-12 |
| 4 | 3.94E-10 | 3.56E-10 | 1.05E-10 | 3.09E-02 | 1.86E-07 | 4.39E-11 | 7.06E-11 | 8.96E-02 | 1.84E-10 |
| 5 | 5.74E-10 | 4.42E-10 | 1.77E-10 | 5.00E-02 | 2.10E-09 | 2.02E-091 | 2.07E-11 | 3.04E-02 | 9.44E-12 |
| 6 | 3.94E-10 | 4.11E-10 | 1.77E-10 | 3.09E-02 | 9.58E-11 | 3.24E-11 | 9.87E-12 | 4.60E-02 | 9.44E-12 |
| 7 | 3.94E-10 | 3.62E-10 | 1.04E-10 | 7.21E-11 | 1.13E-07 | 4.38E-11 | 7.05E-11 | 4.75E-02 | 1.95E-10 |
| 8 | 6.60E-10 | 4.74E-10 | 4.03E-09 | 8.00E-12 | 8.36E-07 | 3.43E-10 | 2.07E-11 | 3.09E-02 | 9.38E-12 |
| 9 | 3.95E-10 | 4.07E-10 | 1.76E-10 | 8.94E-11 | 9.61E-11 | 3.25E-11 | 9.84E-12 | 7.53E-07 | 9.38E-12 |
| 10 | 3.94E-10 | 3.56E-10 | 1.05E-10 | 3.09E-02 | 3.31E-11 | 4.39E-11 | 7.06E-11 | 8.96E-02 | 1.84E-10 |
| 11 | 5.72E-10 | 4.42E-10 | 1.14E-08 | 5.00E-02 | 2.10E-09 | 2.02E-09 | 9.87E-12 | 3.04E-02 | 9.44E-12 |
| 12 | 3.94E-10 | 4.11E-10 | 1.77E-10 | 3.09E-02 | 9.62E-11 | 3.24E-11 | 9.87E-12 | 4.60E-02 | 9.44E-12 |

are listed in Table VII. The proposed twelve formats of the approximate OPF model have slightly different objective solutions due to the approximations of different constraints and the different selection of the constraints. The computational CPU time of the approximate OPF model are listed in Table VIII. Note the computational time for the proposed twelve formats of the approximate OPF model include both optimization model construction time in JuMP and the solver time in IPOPT. Similarly, with the increase of the network scale, the required computational CPU time increases. This is reasonable since the number of model variables and constraints increase. Compared with the computational time of the exact OPF model, when the IPOPT solver converges, the required computational time of the approximate OPF model is more or less the same. The results of the maximum active power approximation gaps of the approximate OPF model are listed in Table IX. Most values are very small or negligible. The results of the maximum reactive power approximation gaps of the approximate OPF model are listed in Table X. Most of these values are larger compared with the values of the maximum active power loss approximation gaps listed in Table IX. A method to reduce the approximation gaps is proposed in the next section.

## IV. APPROXIMATION GAP REDUCTION

To reduce the maximum approximation gap of the reactive power loss for the approximate OPF model, a penalty function based method which adds an penalty term $\xi \sum_l q_{o_l}$ to the original objective function of the OPF model is proposed as:

$$f^{'} = f(p_n) + \xi \sum_l q_{o_l} \quad (5a)$$

Where $\xi > 0$ is the penalty coefficient. In this paper, $\xi = 0.3$ is used. The performance of this method to tighten the approximation gap is shown in Table XI and Table XII. It can be seen that the approximation gaps of the reactive power loss are much smaller than the values listed in Table X. For the maximum approximation gaps of the active power loss, the values are more or less the same compared with results listed in Table IX. Since the maximum approximation gaps of the active power loss are already very small, it is not necessary to reduce them.

## V. CONCLUSIONS

This paper formulates six formats of the exact Mesh-BranchFlow model and six formats of the approximate Mesh-BranchFlow model. All the formats are valid for both radial and mesh power networks. By taking the shunt conductive components of the transmission line Π-model into consideration, a linear amapacity constraint is derived. Based on the formulated Mesh-BranchFlow model and the ampacity constraint, twelve formats of the exact OPF model and twelve formats of the approximate OPF model are proposed. The reason of proposing the approximate OPF model is to give a convex formulation which advantages in finding the global optimal solution using available optimization solvers. The numerical investigations using IEEE test cases prove the accuracy and computational efficiency of all the twelve formats of the exact OPF model. For the twelve formats of the approximate OPF model, non-negligible reactive power loss approximation gaps exist for some test cases if the IPOPT solver is used. To reduce the approximation gaps, this paper proposes to use the penalty function in the objective function of the OPF model. The numerical results show that the approximation gaps of reactive power loss are largely reduced. In this way, the AC-feasibility of the approximate Mesh-BranchFlow model is enhanced.

**Zhao Yuan** is an Assistant Professor and the Head of the Electrical Power Systems Laboratory (EPS-Lab) at the University of Iceland. He worked as a Scientist at the Swiss Federal Institute of Technology Lausanne (EPFL) from 2019 to 2021. Zhao received joint PhD degree from KTH Royal Institute of Technology, Comillas Pontifical University and Delft University of Technology in 2018. He proposed and proved the solution existence-and-uniqueness theorems of the convex optimal power flow model. Zhao co-developed the Energy Management System of the 560kWh/720kVA Battery Energy Storage System (BESS) on EPFL campus and the Smart Grid in Aigle Switzerland. Website: https://english.hi.is/staff/zhaoyuan